\documentclass[aps,prl,amsfonts,amssymb,twocolumn,amsmath,preprintnumbers,nofootinbib,floatfix,
showpacs]{revtex4-1}
\usepackage[dvips]{graphics}
\usepackage{graphicx}
\usepackage{bm}
\begin{document}

\title{Triplet contribution to the Josephson current in the 
nonequilibrium superconductor/ferromagnet/superconductor junction}
\author{I. V. Bobkova}
\affiliation{Institute of Solid State Physics, Chernogolovka,
Moscow reg., 142432 Russia}
\author{A. M. Bobkov}
\affiliation{Institute of Solid State Physics, Chernogolovka,
Moscow reg., 142432 Russia}

\date{\today}

\begin{abstract}
The Josephson current through a long s-wave superconductor/weak ferromagnet/s-wave superconductor
weak link is studied theoretically in the regime of nonequilibrium spin-dependent occupation
of electron states in the ferromagnetic intelayer. While under the considered nonequilibrium
condition the standard supercurrent, carried by the singlet part of current-carrying density of states, is not
modified, the additional supercurrent flowing via the triplet part of the current-carrying density of states
appears. Depending on voltage, controlling the particular form of spin-dependent 
nonequilibrium in the interlayer, this additional current can enhance or reduce the usual current of the singlet
component and also switch the junction between $0$- and $\pi$-states.  
\end{abstract}

\maketitle

New states have been predicted and observed in Josephson weak links in the last years. One of them is the
famous LOFF-state, which was predicted long ago \cite{larkin64,fulde64}. It was not observed in bulk materials,
but can be induced in superconductor/weak ferromagnet/superconductor (SFS) Josephson junction, as it was predicted
theoretically \cite{buzdin82,buzdin92} and observed experimentally \cite{ryazanov01,kontos02,blum02,guichard03}.
The LOFF-state is qualitatively different from the ordinary zero-momentum superconducting state: as a response
to the energy difference between the two spin directions Cooper pair acquires the total momentum
$2Q$ or $-2Q$ inside the ferromagnet. Here $Q \propto h/v_f$, where $h$ is an exchange energy and $v_f$ is
the Fermi velocity, represents an increase (decrease )of the momentum of an electron
with the energetically favorable (unfavorable) spin. Combination of the two possibilities
results in the spatial oscillations of the condensate wave function $\Psi (x)$ in the ferromagnet along the direction
normal to the SF interface. For the singlet Cooper pair $\Psi_{s} (x) \propto \cos(2Qx)$ \cite{demler97}. The same
picture is also valid in the diffusive limit. The only thing we need to add is an extra decay of the condensate
wave function due to scattering. In the regime $h \gg \Delta$, where $\Delta$ is a superconducting order parameter
(OP) in the leads, the decay length is equal to the magnetic coherence length $\xi_F=\sqrt{D/h}$, while 
the oscillation period is given by
$2\pi \xi_F$. Here $D$ is the diffusion constant in the ferromagnet, $\hbar=1$ throughout the paper.
Because of the oscillations the condensate wave function contains nodes where the phase changes by $\pi$. So,
different signs of the OP can occur in the leads, what results in a negative Josephson current response upon a small
increase of the phase. This is the so-called $\pi$-state.

However, manipulating by the phase difference $\chi$ between the OP's in the superconducting leads is not
the only way to realize the $\pi$-state. The net supercurrent that
flows between the two superconductors depends not only on the actual phase difference $\chi$ between the leads, but also
on the occupation of the electron current-carrying states in the weak link. In the clean limit the current is carried by
the Andreev bound states \cite{kulik70,ishii72,bardeen72}. In the diffusive limit the electron trajectories
are not well defined, and Andreev bound states are no longer the appropriate concept to describe the supercurrent.
The energy spectrum of the superconducting correlations is expressed in a so-called supercurrent-
carrying density of states (SCDOS) \cite{volkov95,wilheim98,yip98}. This quantity represents the density of states
weighted by a factor proportional to the current that each state carries in a certain direction.
Positive and negative parts of
SCDOS give energy-dependent contributions
to the supercurrent in the positive and negative direction. The size and direction of the
total supercurrent depends therefore on the occupied fraction of
these states, which is analogous to the occupation of the discrete
Andreev bound states in a ballistic system. That is one can  obtain negative Josephson current response
to small phase differences and, hence, switch the system into $\pi$-state by creating an appropriate nonequilibrium
quasiparticle distribution in the weak link region. This effect was predicted theoretically \cite{volkov95,wilheim98}
and observed experimentally \cite{baselmans99} for a diffusive  SNS junction.

In the present paper we investigate the effect of the nonequilibrium occupation of the current-carrying states
on the Josephson current in SFS junction in the parameter range $\Delta \ll h \ll \varepsilon_f$, where $\varepsilon_f$
is the Fermi energy of the ferromagnet. This regime is relevant to weak ferromagnetic alloys, which were used for
the experimental realization of magnetic $\pi$-junctions. In this particular study we assume the Thouless energy
$\varepsilon_{Th}=D/d^2 \ll h$, that is the interlayer length $d \gg \xi_F$. It is found that a spin-independent nonequilibrium distribution of quasiparticles
in the ferromagnet practically does not affect the Josephson current, that is the described above  mechanism
of the critical current reversal by the non-equilibrium redistribution of supercurrent-carrying states population
is irrelevant in this case. However, the considered system exhibits novel interesting physics upon creating
{\it spin-dependent} nonequilibrium quasiparticle distribution in the weak link.

It is well known that the presence of an exchange field leads to
the formation of the triplet component of the condensate wave
function in the interlayer. In the case of a homogeneous exchange
field, which is considered in the present paper, only the
component with zero spin projection on the field direction $S_z =
0$ is induced. Combining the two pairs with the total momenta $2Q$
and $-2Q$ into the triplet combination we see that the in-plane
($S_z = 0$) triplet condensate wave function component $\Psi_{t}
(x) \propto \sin(2Qx)$, that is oscillates in space with the same
period as the singlet one, but is shifted by $\pi/2$ with respect
to it. Here we do not discuss the other triplet components with
$S_z=\pm 1$, which are typically induced in case of inhomogeneous
magnetization (See \cite{bergeret05} and references therein).
While their influence on the Josephson current under
nonequilibrium conditions is undoubtedly of great interest, they
do not exhibit damped oscillation behavior, which is crucial for
us in the present study.

The proximity-induced triplet component of the condensate function is an odd-frequency quantity \cite{bergeret05}.
Consequently, SCDOS generated by the in-plane triplet component is even function
of quasiparticle energy. The supercurrent can be expressed via the current-carrying density as \cite{yip98}
\begin{equation}
j\propto  \int d \varepsilon N_j (\varepsilon)\varphi(\varepsilon)
\label{supercur}
\enspace ,
\end{equation}
where $\varepsilon$ stands for the quasiparticle energy, $N_j$ denotes SCDOS and
$\varphi(\varepsilon)$ is the distribution function. Since $\varphi(\varepsilon)=\tanh \varepsilon/2T$ in equilibrium,
it is easily seen that in-plane triplet component does not contribute to the supercurrent in this case. It is shown
below that if the distribution function becomes nonequilibrium and spin-dependent, the supercurrent carried by
the triplet component in the ferromagnet is non-zero. It should be noted that this current $j_{tr}$ does not compete with
that one carried by the singlet component $j_s$, but is additional to it. The magnitude of $j_{tr}$ can be of
the same order or even larger than $j_s$. Due to the fact that the singlet and triplet
components have the same oscillation period but shifted in phase by $\pi/2$, $j_{tr}$ can increase the usual supercurrent,
carried by the singlet component, or weaken it, or even reverse the sign of the total supercurrent, thus switching
between $0$ and $\pi$-state. Experimentally the most probable way to realize the spin-dependent nonequilibrium
in the interlayer is to apply a voltage to (or to pass the dissipative current through) a spin-active material. Then
under appropriate conditions even quite small voltages should be enough to switch the system from $0$ to $\pi$-state
and vice versa.

For a quantitative analysis we use the formalism of quasiclassical Green-Keldysh functions in the diffusive limit
\cite{serene83,usadel}.
The fundamental
quantity for diffusive transport is the momentum average of the
quasiclassical Green's function $\check g(x,\varepsilon) =
\langle \check g(\bm p_f, x,\varepsilon) \rangle_{\bm p_f}$. It
is a $8\times8$ matrix form in the product space of Keldysh,
particle-hole and spin variables. Here $x$ is the coordinate
measured along the normal to the junction.

The electric current should be calculated via Keldysh part of the
quasiclassical Green's function. For the plane diffusive junction
the corresponding expression reads as follows
\begin{equation}
j = \frac{-d}{8 \pi^2 e R_F} \int
\limits_{-\infty}^{+\infty} d \varepsilon {\rm Tr}_4
\left[\frac {\tau_0 + \tau_3}{2}
\left(\check g(x, \varepsilon)\partial _x \check
g(x, \varepsilon)\right)^K \right] \label{tok}
,
\end{equation}
where $e$ is the electron charge and $R_F$ stands for the resistance of the ferromagnetic region.
$\left(\check g(x, \varepsilon)\partial_x \check
g(x, \varepsilon)\right)^K$ is a $4\times4$
Keldysh part of the corresponding combination of full Green's
functions. $\tau_i$ and $\sigma_i$ are Pauli matrices in
particle-hole and spin spaces respectively.

It is convenient to express the Keldysh part of the full Green's function via the retarded and advanced components
and the distribution function: $\check g^K=\check g^R \check \varphi-\check \varphi \check g^A$. Here the argument
$(x, \varepsilon)$ of all the functions is omitted for brevity. The distribution function is diagonal in particle-hole
space: $\check \varphi=\hat \varphi (\tau_0+\tau_3)/2+ \sigma_2 \hat {\tilde \varphi} \sigma_2
(\tau_0-\tau_3)/2$. All the matrices denoted by $\hat ~$ are $2 \times 2$ matrices in spin space throughout
the paper. In terms of the distribution function the current (\ref{tok}) takes the form
\begin{eqnarray}
j = \frac{-d}{e R_F} \int
\limits_{-\infty}^{+\infty} \frac{d \varepsilon}{8 \pi^2} {\rm Tr}_2
\left[-\pi^2 \partial _x \hat \varphi - \hat
g^R \partial _x \hat \varphi \hat g^A -  \right. \nonumber \\
-\hat f^R \sigma_2 \partial_x \hat {\tilde \varphi}\sigma_2 \hat {\tilde f}^A +
(\hat g^R \partial_x \hat g^R +\hat f^R \partial_x \hat {\tilde f}^R) \hat \varphi- \nonumber \\
\left.-\hat \varphi (\hat g^A \partial_x \hat g^A +\hat f^A \partial_x \hat {\tilde f}^A) \right]
\label{current_distribution}
.
\end{eqnarray}
We assume that the direction of the exchange field $\bm h$ is spatially homogeneous and choose the quantization axis
along the field. In this case the distribution
function and the normal part $\hat g^{R,A}$ of the Green's function are diagonal matrices in spin space. The anomalous Green's
functions are represented as $\hat f^{R,A}=\hat f_d^{R,A}i \sigma_2$ and $\hat {\tilde f}^{R,A}=
-i\sigma_2\hat {\tilde f}_d^{R,A}$, where $\hat f_d^{R,A}$ and $\hat {\tilde f}_d^{R,A}$ are diagonal in spin space.
That is the last two terms in Eq.~(\ref{current_distribution}) can be rewritten as
$(\hat g^R \partial_x \hat g^R +\hat f^R \partial_x \hat {\tilde f}^R) -\hat g^A \partial_x \hat g^A -
\hat f^A \partial_x \hat {\tilde f}^A)\hat \varphi \propto N_j(\varepsilon) \varphi$. So these two terms
are responsible for the supercurrent carried by SCDOS $N_j(\varepsilon)$ occupied
according to the distribution function $\varphi(\varepsilon)$. The first three terms originate from the direct
diffusion of quasiparticles and are absent in equilibrium.

The retarded and advanced Green's functions are obtained by solving the Usadel equations \cite{usadel}
supplemented with the appropriate boundary conditions at SF interfaces \cite{kupriyanov88}. We assume
the junction to be long, that is $d \gg \xi_F$. This condition allows us to find the solution analytically
even for an arbitrary SF interface transparency and low temperature, that is in the parameter region, where the
equations cannot be linearized with respect to the anomalous Green's function. We start from
the completely incoherent junction (that is consider the left and right SF interfaces separately) and then calculate
the corrections up to the first order of the small parameter $\exp{[-d/\xi_F]}$
to the Green's functions. Within this accuracy the anomalous Green's functions in the vicinity of left (l)
and right (r) SF interfaces take the form
\begin{eqnarray}
f_{d \sigma}^{R,A}=\kappa i \pi \left[ \sinh \Theta_{\sigma}^{R,A}e^{-i\alpha \chi/2}+
4 \Sigma_{2,\sigma}^{R,A}(x) e^{i\alpha \chi /2} \right]
\enspace , \nonumber \\
\tilde f_{d \sigma}^{R,A}=-f_{d \sigma}^{R,A}(\chi \to -\chi)\enspace .
\qquad \qquad
\label{f}
\end{eqnarray}
Here $\sigma=\uparrow ,\downarrow $ is the electron spin projection on the quantization axis, $\kappa=+1(-1)$ corresponds
to the retarded (advanced) functions and $\alpha =+1(-1)$ in the vicinity of the left (right) SF interface.
The first
term represents the anomalous Green's function at the ferromagnetic side of the isolated SF boundary with
\begin{equation}
\sinh \Theta_\sigma^{R,A}=\frac{4 \Sigma _{1,\sigma}^{R,A}(x)
(1+{\Sigma _{1,\sigma}^{R,A}}^2(x))}
{(1-{\Sigma _{1,\sigma}^{R,A}}^2(x))^2}
\label{sh_Theta}
\enspace ,
\end{equation}
while the second one is the first order correction, originated from the anomalous Green's function
extended from the other SF interface.
\begin{equation}
\Sigma _{1(2),\sigma}^{R,A}(x)=K_\sigma ^{R,A}e^{\displaystyle -(\frac{d}{2}+(-)\alpha x)(1+i\kappa \sigma)/\xi_F}
\label{Sigma12}
\enspace ,
\end{equation}
where $K_\sigma^{R,A}$ is determined by the equation
\begin{eqnarray}
(1+i\kappa \sigma)K_\sigma^{R,A}(1-{K_\sigma^{R,A}}^2)=
\frac{R_F \xi_F}{4R_g d}\left[ \sinh \Theta_s^{R,A}(1+ \right. \nonumber \\
\left. +{K_\sigma^{R,A}}^2+{K_\sigma^{R,A}}^4)-
\cosh \Theta_s^{R,A} 4 K_\sigma^{R,A}(1+{K_\sigma^{R,A}}^2)\right]
\label{K}
\enspace .
\end{eqnarray}
$R_g$ stands for the resistance of each SF interface, which are
supposed to be identical. $\sinh \Theta_s^{R,A}$ and $\cosh
\Theta_s^{R,A}$ originated from the Green's functions in the
superconducting leads, which are assumed to be homogeneous and
equilibrium. In this case
\begin{eqnarray}
\cosh \Theta_s^{R,A}=\frac{-\kappa
i\varepsilon}{\sqrt{\Delta^2-(\varepsilon+\kappa i\delta)^2}}
\nonumber \\
\sinh \Theta_s^{R,A}=\frac{-\kappa i
\Delta}{\sqrt{\Delta^2-(\varepsilon+\kappa i \delta)^2}}
\label{bulk_Greens_functions} \enspace ,
\end{eqnarray}
where $\delta$ is a positive infinitesimal.

The nonequilibrium distribution function in the interlayer is
supposed to be created by applying a voltage along the $y$
direction between two additional electrodes $N_b$ and $N_t$, which
are attached to the central part of the interlayer. It
is assumed that the conductances of $N_b$F and $N_t$F interfaces
are spin-dependent and equal to ${g_b}_\sigma$ and ${g_t}_\sigma$,
respectively. The voltage ${V_t}_\sigma$ (${V_b}_\sigma$) between the superconducting leads
and $N_t$ ($N_b$) electrode can also be spin-dependent. It can be realized, for example,
by taking the electrodes $N_b$ and $N_t$ to be ferromagnetic and passing the dissipative current 
between them. Then
the expression for the distribution function in the middle of the interlayer $\varphi^{(0)}_\sigma$ to the zero
order of the parameter $\exp{[-d/\xi_F]}$ (disregarding the proximity effect) 
takes the form
\begin{eqnarray}
\varphi^{(0)}_\sigma=\frac{\tanh \frac{\varepsilon-e{V_t}_\sigma}{2T}{g_t}_\sigma(\sigma_F+d_y {g_b}_\sigma)+
}{\sigma_F({g_t}_\sigma+{g_b}_\sigma)+} \nonumber \\
\frac{+\tanh \frac{\varepsilon-e{V_b}_\sigma}{2T}{g_b}_\sigma (\sigma_F+d_y {g_t}_\sigma)}{+2d_y{g_t}_\sigma{g_b}_\sigma}
\label{phi_N}
\enspace ,
\end{eqnarray}
where $T$ is the temperature, $\sigma_f$ is the conductivity of F interlayer and $d_y$ is its width in the $y$ direction.
$\tilde {\varphi}^{(0)}_\sigma$ is
connected to $\varphi^{(0)}_\sigma$ by the relation $\tilde {\varphi}^{(0)}_{\uparrow ,\downarrow}(\varepsilon)=
-\varphi^{(0)}_{\downarrow ,\uparrow}(-\varepsilon)$.

However, for the considered problem it is not enough to only calculate the distribution function without taking into
account the proximity effect in the interlayer. It should be obtain up to the first order of the parameter 
$\exp{[-d/\xi_F]}$. For this purpose we solve the kinetic equation for the distribution function,
which can be easily
derived from the Keldysh part of the Usadel equation . The
boundary conditions to the kinetic equation are also obtained from
the Keldysh part of the general Kypriyanov-Lukichev boundary
conditions \cite{kupriyanov88}.

Substituting the Green's functions (\ref{f}) and the distribution functions into Eq.~(\ref{current_distribution})
we find the expression for Josephson current, which in the limit of low temperatures $T \ll |eV_{t,b}|$ and
for $|{eV_{t,b}}_{\uparrow ,\downarrow}|<\Delta$ (the last condition allows us to avoid flowing of the dissipative
current from the additional electrodes to the superconducting leads) takes the form 
\begin{eqnarray}
j=\frac{-\sin \chi}{2eR_g}\int d \varepsilon \sum \limits_\sigma 
\left[ (\varphi^{(0)}_\sigma+\tilde {\varphi}^{(0)}_\sigma) \times \right. \nonumber \\
\left. \times {\rm Im} (\Sigma_{2,\sigma}^R(-\alpha \frac{d}{2})\sinh 
\Theta_s^R) \right]
\label{current_general}
\enspace .
\end{eqnarray}
Eq.~(\ref{current_general}) is the central result of our paper. While it is valid for arbitrary SF interface
transparency, we concentrate on the discussion of the tunnel limit $\tilde g \equiv R_F\xi_F/R_gd \ll 1$, where Eq.~(\ref{K})
can be easily solved and the integral over energy can be calculated analytically. For the analytical analysis 
we choose the most simple form for the distribution function Eq.~(\ref{phi_N}) by setting ${g_t}_\sigma=0$ and
${V_b}_\downarrow = 0$. Then $\varphi^{(0)}_\uparrow = \tanh[(\varepsilon-e{V_b}_\uparrow)/2T]$, while 
$\varphi^{(0)}_\downarrow = \tanh[\varepsilon/2T]$ and we obtain for the Josephson current the following result     
\begin{eqnarray}
j=\frac{R_F \xi_F \sin \chi}{4eR_g^2 d}\Delta e^{-d/\xi_F}\left[ 
\sqrt 2 \pi \cos(\frac{d}{\xi_F}+\frac{\pi}{4})+\right. \nonumber \\
\left. \frac{1}{\sqrt 2}\log \left| \frac{\Delta+e{V_b}_\uparrow}{\Delta-e{V_b}_\uparrow} \right|
\sin (\frac{d}{\xi_F}+\frac{\pi}{4}) \right]
\label{current_tunnel}
\enspace .
\end{eqnarray}
The first term in Eq.~(\ref{current_tunnel}) represents the part of the supercurrent carried by the singlet
component of SCDOS. Under the conditions $T \ll |eV_{t,b}|$ and
$|{eV_{t,b}}_{\uparrow ,\downarrow}|<\Delta$ it is not affected by the fact that the distribution function 
is nonequilibrium. The reason is that for a long junction and $h \gg \Delta$ the singlet part of SCDOS is concentrated in the narrow energy intervals around $\varepsilon=\pm \Delta$, where the distribution
function does not differ from the equilibrium value. The second term $j_{tr}$ is caused by the triplet component and for
qualitative understanding can be rewritten as
\begin{eqnarray}
j_{tr} \propto \sum \limits_\sigma \int \limits_{-\Delta}^{\Delta} 
d\varepsilon (N_j)_{tr,\sigma}\left[\varphi^{(0)}_\sigma+ 
\frac 12 (\tilde {\varphi}^{(0)}_\sigma-\varphi^{(0)}_\sigma)\right]
\label{current_qualitative}
,
\end{eqnarray}
where the triplet component of SCDOS $(N_j)_{tr,\sigma} \propto \sigma e^{-d/\xi_F}
\frac{\Delta^2}{\Delta^2-\varepsilon^2}\sin[d/\xi_F+\pi/4]$. The first term in 
Eq.~(\ref{current_qualitative}) can be interpreted as the part of the supercurrent carried by the triplet
component of SCDOS. Contrary to the singlet component of SCDOS, which is an odd function of energy, $(N_j)_{tr,\sigma}(\varepsilon)=
(N_j)_{tr,\sigma}(-\varepsilon)$ and, consequently, does not contribute to supercurrent in equilibrium.
It is worth to note here that the spin-independent distribution function also does not give rise to the flow
of the triplet component, because the total combination of the distribution functions in the square brakets
is odd function of energy in this case.

It is well known that the spin supercurrent can not flow through the singlet superconducting lead. 
On the other hand, it is seen that for our choice of the distribution function the part of the supercurrent 
determined by the first term in Eq.~(\ref{current_qualitative}) is carried entirely by spin-up electrons.
However, the constant gradient of the distribution function $\propto (N_j)_{tr}$ appears
in the interlayer and cause the diffusive flow of electrons, which compensate the spin current. This flow
originates from the first three terms in Eq.(\ref{current_distribution}) and is described by the second term
in Eq.~(\ref{current_qualitative}).

It can be easily seen from Eq.~(\ref{current_tunnel}) that by manipulating ${V_b}_\uparrow$ one can
enhance or reduce the Josephson current, or to switch the junction between $0$- and $\pi$-states.

\begin{figure}[!tbh]
   \begin{minipage}[b]{.5\linewidth}
     \centerline{\includegraphics[clip=true,width=1.71in]{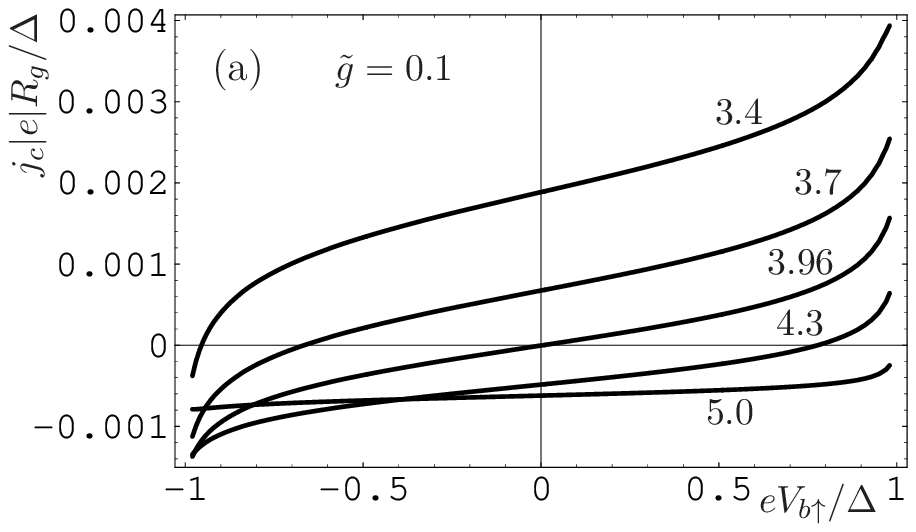}}
     \end{minipage}\hfill
    \begin{minipage}[b]{.5\linewidth}
   \centerline{\includegraphics[clip=true,width=1.6in]{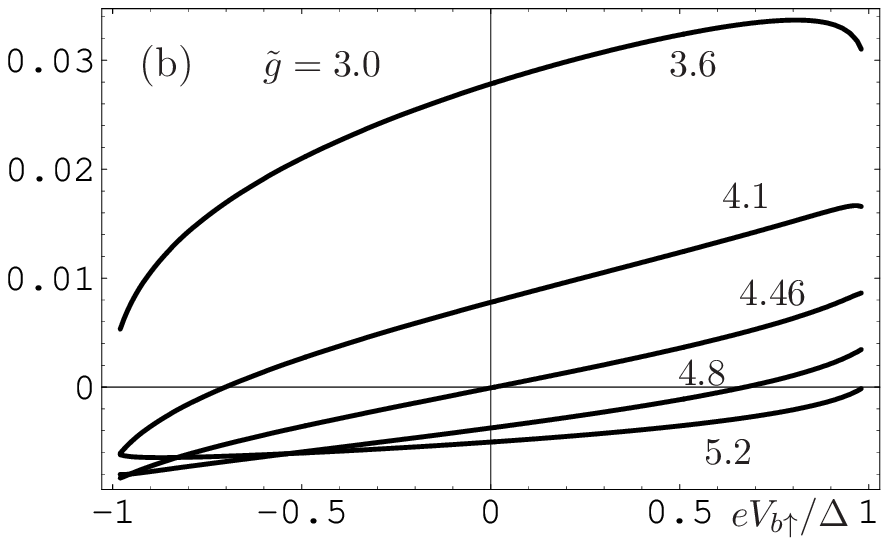}}
  \end{minipage}
   \caption{The critical Josephson current as a function of $e{V_b}_\uparrow/\Delta$. The current is measured in units
$j_c|e|R_g/\Delta$. The different curves correspond to different lengths $d$ of the junction, which are measured
in units of $\xi_F$.}
\label{current}
\end{figure}

To calculate the Josephson current for arbitrary transparency of SF interfaces one needs to solve Eq.~(\ref{K})
numerically and make use of Eq.~(\ref{current_general}). The corresponding curves are plotted in Fig.~\ref{current}.
Panel (a) shows the current for low enough dimensionless conductance of SF interface $\tilde g=0.1$, while 
panel (b) represents the case of highly transparent interface $\tilde g=3.0$. Different curves correspond
to different length $d$ of the junction. It is seen that the discussed above tunnel 
limit qualitatively captures the essential physics. In dependence on ${V_b}_\uparrow$ the current can be enhanced or reduced
with respect to its value at ${V_b}_\uparrow=0$. If the length of the equilibrium junction is not far from the
$0$-$\pi$ transition, then small enough voltage can switch between the states.

In conclusion, we have studied the effects of nonequilibrium spin-dependent electron distribution in the 
weakly ferromagnetic interlayer on the Josephson current through SFS junction. It is shown that the
nonequilibrium spin-dependent electron distribution gives rise to the supercurrent carried by the triplet
component of SCDOS. Depending on voltage, controlling the particular form of spin-dependent 
nonequilibrium in the interlayer, this additional current can enhance or reduce the usual current of the singlet
component and also switch the junction between $0$- and $\pi$-states. The principal difference of the considered effect
from already discussed in the literature voltage driven SNS transistor \cite{volkov95,wilheim98,yip98,
baselmans99} is that in our case not the population of the singlet part of SCDOS
is redistributed leading to diminishing of the current and reversing its sign, but the additional triplet 
component begins to contribute to supercurrent.

We gratefully acknowledge the discussions with V. Ryazanov. The
support by RFBR Grant 09-02-00779 and the programs of Physical Science Division
of RAS is acknowledged. A.M.B. was also supported by the Russian
Science Support Foundation.


\end{document}